\documentclass[12pt,a4paper]{article}

\begin{document}

\title{Nonleptonic kaon decays: theory vs. experiment}

\author{C.Cheshkov\\{\it\small Joint Institute for Nuclear Research, Dubna, Russian Federation}}

\maketitle

\begin{abstract}
A review of the present experimental status of the rates and the Dalitz-plot
slope parameters in CP conserving \( K\rightarrow 2\pi  \) and \( K\rightarrow 3\pi  \)
decays is given. The corresponding isospin amplitudes have been determined by
a common fit to the recent experimental data and have been used as an input
to a consequent fit based on the constraints from \( O\left( p^{4}\right)  \)
chiral lagrangian. It has been found that the constraints of the chiral fit
are well satisfied by the experimental data, allowing to estimate the weak coupling
constants and to predict the quadratic amplitudes in \( K\rightarrow 3\pi  \)
decays and the one-loop strong rescattering phases. The consistency of the obtained
weak coupling constants with weak resonance models has been also considered.
\end{abstract}

\pagebreak

\section{Introduction}

Quantum Chromodynamics (QCD) is the established theory for describing strong
quark-quark interactions. However, the theoretical study of kaon decays is hindered
by the nonperturbative behavior of QCD in the low-energy domain. One of the
approaches successfully applied to the study of low-energy processes is chiral
perturbation theory (ChPT) \cite{Pich,Gasser}. It is based on the chiral symmetry
properties of QCD and the concept of the effective field theory. An important
point in the application of ChPT is the existence of coupling constants which
can not be calculated from first principles and therefore must be extracted
phenomenologically from the experimental data.

In the present paper we report an update of the analysis of CP conserving \( K\rightarrow 2\pi  \)
and \( K\rightarrow 3\pi  \) decays, performed in \cite{Kambor}. The update
is caused by necessity to take into account the recent measurements of \( \Gamma \left( K_{S}\rightarrow \pi ^{+}\pi ^{-}\pi ^{0}\right)  \)
and the Dalitz plot parameters in \( K\rightarrow 3\pi  \) decays, carried
out by the CPLEAR \cite{CPLEAR}, the HYPERON \cite{HYPERON} and the NA48 \cite{Rainer}
collaborations.

In section \ref{phenomen} we present the results from a phenomenological fit
of \( K\rightarrow 2\pi  \) and \( K\rightarrow 3\pi  \) isospin amplitudes
to the recent experimental data. In section \ref{chpttheory} we consider the
estimation of the isospin amplitudes in terms of \( O\left( p^{4}\right)  \)
chiral lagrangian and summarize the results from a fit of the weak coupling
constants. In section \ref{resonance} we briefly discuss the consistency of
the obtained results with weak resonance models. Section \ref{conclusions}
contains some concluding remarks.

\section{\label{phenomen}Phenomenological fit}

Using the usual isospin decomposition CP conserving \( K\rightarrow 2\pi  \)
amplitudes can be expressed in the following way\footnote{%
assuming \( \Delta I\leq 3/2 \)
}:

\begin{equation}
\label{k2piisospin}
\begin{array}{rcl}
A\left( K^{0}\rightarrow \pi ^{+}\pi ^{-}\right)  & = & -\frac{1}{\sqrt{3}}A_{0}+\frac{1}{\sqrt{6}}A_{2}\\
A\left( K^{0}\rightarrow \pi ^{0}\pi ^{0}\right)  & = & \frac{1}{\sqrt{3}}A_{0}+\sqrt{\frac{2}{3}}A_{2}\\
A\left( K^{+}\rightarrow \pi ^{+}\pi ^{0}\right)  & = & \frac{\sqrt{3}}{2}A_{2}
\end{array}
\end{equation}
where \( A_{0}=ia_{1/2}e^{i\delta _{0}} \) and \( A_{2}=-ia_{3/2}e^{i\delta _{2}} \)
represent transitions with \( \Delta I=1/2,\: 3/2 \), respectively.

Due to the small available phase space in \( K\rightarrow 3\pi  \) decays their
amplitudes can be expanded in powers of Dalitz plot variables \( u \) and \( v \):

{\footnotesize \begin{equation}
\label{k3piisospin}
\begin{array}{rcl}
A\left( K_{L}\rightarrow \pi ^{+}\pi ^{-}\pi ^{0}\right)  & = & \left( \alpha _{1}+\alpha _{3}\right) -\left( \beta _{1}+\beta _{3}\right) u\\
 & + & \left( \zeta _{1}-2\zeta _{3}\right) \left( u^{2}+v^{2}/3\right) +\left( \xi _{1}-2\xi _{3}\right) \left( u^{2}-v^{2}/3\right) \\
A\left( K_{L}\rightarrow \pi ^{0}\pi ^{0}\pi ^{0}\right)  & = & -3\left( \alpha _{1}+\alpha _{3}\right) -3\left( \zeta _{1}-2\zeta _{3}\right) \left( u^{2}+v^{2}/3\right) \\
A\left( K^{+}\rightarrow \pi ^{+}\pi ^{+}\pi ^{-}\right)  & = & \left( 2\alpha _{1}-\alpha _{3}\right) +\left( \beta _{1}-\beta _{3}/2+\sqrt{3}\gamma _{3}\right) u\\
 & + & 2\left( \zeta _{1}+\zeta _{3}\right) \left( u^{2}+v^{2}/3\right) -\left( \xi _{1}+\xi _{3}-\xi ^{'}_{3}\right) \left( u^{2}-v^{2}/3\right) \\
A\left( K^{+}\rightarrow \pi ^{0}\pi ^{0}\pi ^{+}\right)  & = & -\left( \alpha _{1}-\alpha _{3}/2\right) +\left( \beta _{1}-\beta _{3}/2-\sqrt{3}\gamma _{3}\right) u\\
 & - & \left( \zeta _{1}+\zeta _{3}\right) \left( u^{2}+v^{2}/3\right) -\left( \xi _{1}+\xi _{3}+\xi ^{'}_{3}\right) \left( u^{2}-v^{2}/3\right) \\
A\left( K_{S}\rightarrow \pi ^{+}\pi ^{-}\pi ^{0}\right)  & = & \left( 2/\sqrt{3}\right) \gamma _{3}v-\left( 4/3\right) \xi ^{'}_{3}uv
\end{array}
\end{equation}
}where

\begin{equation}
\label{dalitzvars}
u=\frac{s_{3}-s_{0}}{M^{2}_{\pi ^{\pm }}},\; v=\frac{s_{2}-s_{1}}{M^{2}_{\pi ^{\pm }}},\; s_{i}=\left( P_{K}-P_{\pi _{i}}\right) ^{2},\; s_{0}=\frac{\left( s_{1}+s_{2}+s_{3}\right) }{3}
\end{equation}
and \( P_{K} \), \( P_{\pi _{i}} \) are the four-momenta of the kaon and \( i \)-th\footnote{%
index \( 3 \) refers to the {}``odd{}'' pion
} pion. The subscripts \( 1 \) and \( 3 \) in (\ref{k3piisospin}) correspond
to \( \Delta I=1/2,\: 3/2 \) transitions, respectively. Neglecting the strong
rescattering final state phases and assuming CP invariance all the isospin amplitudes
\( \alpha _{1} \), \( \alpha _{3} \), \( \beta _{1} \), \( \beta _{3} \),
\( \gamma _{3} \), \( \zeta _{1} \), \( \zeta _{3} \), \( \xi _{1} \), \( \xi _{3} \)
and \( \xi ^{'}_{3} \) are real.

Until now the experimentally measured event distributions for CP conserving
\( K\rightarrow 3\pi  \) decays have been analyzed in terms of the Dalitz plot
slopes \( g \), \( h \), \( k \):

\begin{equation}
\label{dalitzplot}
\left| A\left( K\rightarrow 3\pi \right) \right| ^{2}\sim 1+gu+hu^{2}+kv^{2}\; .
\end{equation}
The recent experimental data \cite{PDG} on the decay widths and the Dalitz
plot slopes are reported in Table \ref{expslopes}\footnote{%
We have included the value of \( (-6.1\pm 0.9_{stat}\pm 0.5_{syst})\times 10^{-3} \)
for the quadratic Dalitz plot slope parameter \( h \) in the \( K_{L}\rightarrow 3\pi ^{0} \)
decay, measured by the NA48 collaboration \cite{Rainer}
}.

The phenomenological fit of \( K\rightarrow 3\pi  \) isospin amplitudes to
the measured decay widths and Dalitz plot slopes was introduced in \cite{Devlin}
and was significantly improved in \cite{Kambor} by including the \( \Delta I=3/2 \)
quadratic amplitudes \( \zeta _{3} \), \( \xi _{3} \) and \( \xi ^{'}_{3} \).
We have redone the fit by adding the measurements of \( \Gamma \left( K_{s}\rightarrow \pi ^{+}\pi ^{-}\pi ^{0}\right)  \)
and the quadratic Dalitz plot slope \( k \) in \( K^{+}\rightarrow \pi ^{0}\pi ^{0}\pi ^{+} \).
The amplitudes \( a_{1/2} \), \( a_{3/2} \) and the relative phase \( \left( \delta _{2}-\delta _{0}\right)  \)
between them have been evaluated using only the measured \( K\rightarrow 2\pi  \)
decay widths. The results for the isospin amplitudes are shown in the first
column of Table \ref{chptfit}. Due to small isospin breaking effects the obtained
value of \( -64.6^{\circ } \) for the relative phase between \( a_{3/2} \)
and \( a_{1/2} \) are not reliable and has not been used in the further analysis.
The relatively high value of \( \chi ^{2} \) of the fit is due entirely to
an inconsistency of the precisely measured \( K_{L}\rightarrow 3\pi  \) and
\( K^{\pm }\rightarrow 3\pi  \) decay widths. Contrariwise, all the measured
Dalitz plot slopes and \( \Gamma \left( K_{S}\rightarrow \pi ^{+}\pi ^{-}\pi ^{0}\right)  \)
seems to be quite consistent.

We have also tried to extract the strong rescattering phase of the amplitudes
\( \beta  \) from the available data. In order to do this one have to replace
the real amplitudes \( \beta  \) in (\ref{k3piisospin}) with

\begin{equation}
\label{betaredefinition}
\beta \rightarrow \beta e^{i\delta _{\beta }}\approx \beta (1+i\delta _{\beta })\: ,
\end{equation}
where \( \delta _{\beta } \) is real \cite{Belkov}. Since the phases of the
constant and quadratic amplitudes are expected to be considerably smaller than
\( \delta _{\beta } \) \cite{Ambrosio}, they have been neglected. The phase
\( \delta _{\beta } \) has been introduced into the phenomenological fitting
procedure as a freely varying parameter. As a result we have obtained \( \left| \delta _{\beta }\right| <0.3 \)
(\( 68\% \) CL) while the isospin amplitudes have remained almost unchanged.

\section{\label{chpttheory}Theoretical estimations within \protect\( O\left( p^{4}\right) \protect \)
ChPT}

To the lowest order \( O\left( p^{2}\right)  \), CP conserving weak chiral
Lagrangian is a sum of \( \left( 8_{L},1_{R}\right)  \) and \( \left( 27_{L},1_{R}\right)  \)
operators \cite{Kambor2}:

\begin{equation}
\label{op2lagrangian}
\mathcal{L}^{\left( 2\right) }_{W}=c_{2}Tr\lambda _{6}L_{\mu }L^{\mu }+c_{3}t_{ik}^{jl}\left( TrQ_{j}^{i}L_{\mu }\right) \left( TrQ_{l}^{k}L^{\mu }\right) \: ,
\end{equation}
where \( \left( Q_{j}^{i}\right) _{ab}=\delta _{ib}\delta _{ja} \) are the
projectors in flavor space, \( \lambda _{6}=Q^{2}_{3}+Q^{3}_{2} \), \( t_{ik}^{jl} \)
are defined to select the \( 27 \)-plet part of the interaction, \( L_{\mu }=iU^{\dagger }\partial _{\mu }U \)
is the left-handed weak current, \( U=e^{\left( i\Phi /F_{\pi }\right) } \),
\( F_{\pi }=92.4 \) \( MeV \) is the pion decay constant and \( \Phi  \)
is the pseudoscalar meson \( SU(3) \) matrix:

\begin{equation}
\label{mesonfield}
\Phi =\sqrt{2}\left( \begin{array}{ccc}
\frac{\pi ^{0}}{\sqrt{2}}+\frac{\eta }{\sqrt{6}} & -\pi ^{+} & -K^{+}\\
\pi ^{-} & -\frac{\pi ^{0}}{\sqrt{2}}+\frac{\eta }{\sqrt{6}} & -K^{0}\\
K^{-} & -\overline{K}^{0} & -\frac{2\eta }{\sqrt{6}}
\end{array}\right) \: .
\end{equation}
The lowest order chiral Lagrangian corresponds to {}``tree{}'' diagrams which
contain one weak vertex and gives the following contributions to the \( \Delta I=1/2 \)
isospin amplitudes \cite{Kambor}:

\begin{equation}
\label{op2deltai12}
\begin{array}{rcl}
\Re eA_{0} & = & \frac{\sqrt{6}}{F^{2}_{\pi }F_{K}}\left( M^{2}_{K}-M^{2}_{\pi }\right) \left( c_{2}-\frac{2}{3}c_{3}\right) \\
\Re e\alpha _{1} & = & \frac{1}{3F^{3}_{\pi }F_{K}}M^{2}_{K}\left( c_{2}-\frac{2}{3}c_{3}\right) \\
\Re e\beta _{1} & = & \frac{-1}{F^{3}_{\pi }F_{K}}M^{2}_{\pi }\left( c_{2}-\frac{2}{3}c_{3}\right) 
\end{array}
\end{equation}
and to the \( \Delta I=3/2 \) ones:

\begin{equation}
\label{op2deltai32}
\begin{array}{rcl}
\Re eA_{2} & = & \frac{-20}{\sqrt{3}F^{2}_{\pi }F_{K}}\left( M^{2}_{K}-M^{2}_{\pi }\right) c_{3}\\
\Re e\alpha _{3} & = & \frac{20}{9F^{3}_{\pi }F_{K}}M^{2}_{K}c_{3}\\
\Re e\beta _{3} & = & \frac{5}{3F^{3}_{\pi }F_{K}}\frac{M^{2}_{\pi }\left( 5M^{2}_{K}-14M^{2}_{\pi }\right) }{\left( M^{2}_{K}-M^{2}_{\pi }\right) }c_{3}\\
\Re e\gamma _{3} & = & \frac{-15}{2\sqrt{3}F^{3}_{\pi }F_{K}}\frac{M^{2}_{\pi }\left( 3M^{2}_{K}-2M^{2}_{\pi }\right) }{\left( M^{2}_{K}-M^{2}_{\pi }\right) }c_{3}
\end{array}
\end{equation}
Thus the coupling constants \( c_{2} \) and \( c_{3} \) of the \( \Delta I=1/2,\: 3/2 \)
operators are determined by the two well-measured \( K\rightarrow 2\pi  \)
isospin amplitudes \( a_{1/2} \) and \( a_{3/2} \). \( SU(3) \) symmetry
breaking is taken into account by fixing the kaon decay constant \( F_{K} \)
to its physical value of \( 113 \) \( MeV \).

Now lets consider the corrections to the isospin amplitudes due to the one-loop
contributions from \( \mathcal{L}^{\left( 2\right) }_{S}\times \mathcal{L}^{\left( 2\right) }_{W} \)
and the tree-level contributions from both \( O\left( p^{4}\right)  \) weak
and strong chiral lagrangians. Following the analysis in \cite{Kambor,Kambor3},
we have factorized \( K\rightarrow 2\pi  \) and \( K\rightarrow 3\pi  \) isospin
amplitudes in the following way:

\begin{equation}
\label{factorize1}
A_{i}=A_{i}^{\left( 2\right) }+A_{i}^{\left( 4\right) }\: ,
\end{equation}
where \( A^{\left( 2\right) }_{i} \) are the lowest order values defined by
equations (\ref{op2deltai12}), (\ref{op2deltai32}) and \( A^{\left( 4\right) }_{i} \)
are the corresponding next-to-lowest corrections. Furthermore, the latter can
be decompressed as:

\begin{equation}
\label{factorize2}
A^{\left( 4\right) }_{i}=A^{loop}_{i}\left( \mu \right) +A^{wk.ct.}_{i}\left( \mu \right) +A^{st.ct.}_{i}\left( \mu \right) \: ,
\end{equation}
where \( A^{loop}_{i}\left( \mu \right)  \), \( A^{wk.ct.}_{i}\left( \mu \right)  \),
\( A^{st.ct.}_{i}\left( \mu \right)  \) are the one-loop contributions and
the tree-level contributions from the counterterms of \( O\left( p^{4}\right)  \)
weak and strong lagrangians, respectively. The parameter \( \mu  \) is the
renormalization scale on which, in principle, \( A^{\left( 4\right) }_{i} \)
must be independent.

Since the one-loop contributions \( A^{loop}_{i}\left( \mu \right)  \) arise
from the lowest order weak lagrangian, they are proportional to the coupling
constants \( c_{2} \) and \( c_{3} \) and are conveniently expressed in terms
of the so-called reduced amplitudes \( a^{\left( 8\right) }_{i} \) and \( a^{\left( 27\right) }_{i} \):

\begin{equation}
\label{reducedloop}
A_{i}^{loop}=\left( c_{2}/F^{2}_{\pi }\right) a_{i}^{\left( 8\right) }+\left( c_{3}/F^{2}_{\pi }\right) a^{\left( 27\right) }_{i}\: .
\end{equation}
For the calculation of the one-loop contributions we have used the values of
the reduced amplitudes for \( \mu =M_{\eta } \) obtained in \cite{Kambor}
and presented in Table \ref{loop}.

As it was shown in \cite{Kambor,Kambor3}, without external fields and neglecting
corrections of order \( M^{2}_{\pi }/M^{2}_{K} \) the contributions from weak
counterterms can be expressed in terms of only \( 7 \) linear combinations\footnote{%
the explicit form of \( K_{i} \) can be found in \cite{Kambor}
} \( K_{i} \), \( i=(1\div 7) \), of weak coupling constants. On the other
side, the contributions from the counterterms of the strong lagrangian are described
by \( 5 \) out of \( 10 \) well-known coupling constants \( L_{i} \), \( i=(1\div 5) \)
(Table \ref{li}) \cite{Gasser,Gasser2}. The counterterm contributions to the
isospin amplitudes are:

{\small \begin{equation}
\label{counterterms}
\begin{array}{rcl}
\Re eA_{0} & = & \frac{-2\sqrt{2}}{3\sqrt{3}F^{2}_{\pi }F_{K}}M^{4}_{K}\left( K_{1}+K_{4}\right) \\
\Re eA_{2} & = & \frac{-20}{3\sqrt{3}F^{2}_{\pi }F_{K}}M^{4}_{K}K_{4}\\
\Re e\alpha _{1} & = & \frac{-2}{27F^{3}_{\pi }F_{K}}M^{4}_{K}\left[ \left( K_{1}+K_{4}-K_{2}+2K_{5}\right) \right. \\
 & + & \frac{4\left( 3c_{2}-2c_{3}\right) }{F^{2}_{\pi }}\left. \left( 4L_{1}+4L_{2}+2L_{3}\right) \right] \\
\Re e\alpha _{3} & = & \frac{20}{27F^{3}_{\pi }F_{K}}M^{4}_{K}\left[ \left( K_{4}+2K_{5}\right) \right. \\
 & - & \frac{8c_{3}}{F^{2}_{\pi }}\left. \left( 4L_{1}+4L_{2}+2L_{3}\right) \right] \\
\Re e\beta _{1} & = & \frac{-1}{9F^{3}_{\pi }F_{K}}M^{2}_{K}M^{2}_{\pi }\left[ \left( -2K_{1}-2K_{4}+K_{3}\right. \right. +\frac{16}{27}K_{5}+\frac{35}{27}K_{6}-\frac{13}{9}\left. K_{7}\right) \\
 & + & \frac{8\left( 3c_{2}-2c_{3}\right) }{F^{2}_{\pi }}\left. \left( -2L_{1}+L_{2}-L_{3}+12L_{4}\right) \right] \\
\Re e\beta _{3} & = & \frac{5}{18F^{3}_{\pi }F_{K}}\frac{M^{4}_{K}M^{2}_{\pi }}{\left( M^{2}_{K}-M^{2}_{\pi }\right) }\left[ \left( 10K_{4}+4K_{6}+3K_{7}\right) \right. \\
 & + & \frac{64c_{3}}{F^{2}_{\pi }}\left. \left( 2L_{1}-L_{2}+L_{3}-12L_{4}\right) \right] \\
\Re e\gamma _{3} & = & \frac{-5}{4\sqrt{3}F^{3}_{\pi }F_{K}}\frac{M^{4}_{K}M^{2}_{\pi }}{\left( M^{2}_{K}-M^{2}_{\pi }\right) }\left[ \left( 6K_{4}+K_{7}\right) \right. \\
 & + & \frac{8c_{3}}{F^{2}_{K}}\frac{M^{2}_{\pi }}{M^{2}_{K}}\left. \left( -L_{3}+12L_{4}+6L_{5}\right) \right] \\
\Re e\zeta _{1} & = & \frac{-1}{6F^{3}_{\pi }F_{K}}M^{4}_{\pi }\left[ \left( K_{2}-2K_{5}\right) \right. \\
 & - & \frac{8\left( 3c_{2}-2c_{3}\right) }{F^{2}_{\pi }}\left. \left( 2L_{1}+2L_{2}+L_{3}\right) \right] \\
\Re e\zeta _{3} & = & \frac{5}{3F^{3}_{\pi }F_{K}}M^{4}_{\pi }\left[ K_{5}\right. \\
 & - & \frac{8c_{3}}{F^{2}_{\pi }}\left. \left( 2L_{1}+2L_{2}+L_{3}\right) \right] \\
\Re e\xi _{1} & = & \frac{-1}{6F^{3}_{\pi }F_{K}}M^{4}_{\pi }\left[ \left( K_{3}\right. \right. +\frac{16}{27}K_{5}+\frac{35}{27}K_{6}-\frac{13}{9}\left. K_{7}\right) \\
 & - & \frac{8\left( 3c_{2}-2c_{3}\right) }{F^{2}_{\pi }}\left. \left( 2L_{1}-L_{2}+L_{3}\right) \right] \\
\Re e\xi _{3} & = & \frac{-5}{24F^{3}_{\pi }F_{K}}M^{4}_{\pi }\left[ \left( 4K_{6}+3K_{7}\right) \right. \\
 & + & \frac{64c_{3}}{F^{2}_{\pi }}\left. \left( 2L_{1}-L_{2}+L_{3}\right) \right] \\
\Re e\xi ^{'}_{3} & = & \frac{15}{8F^{3}_{\pi }F_{K}}M^{4}_{\pi }\left[ K_{7}\right. \\
 & - & \frac{8c_{3}}{F^{2}_{K}}\frac{M^{2}_{\pi }}{M^{2}_{K}}\left. L_{3}\right] \; .
\end{array}
\end{equation}
}Using the twelve isospin amplitudes in the first column of Table \ref{chptfit}
and requiring that all the nine relative phases of \( K\rightarrow 3\pi  \)
amplitudes be zero within errors of \( 15^{\circ } \) (compatible with the
upper bound of \( \left| \delta _{\beta }\right|  \)), we have fitted the coupling
constants \( c_{2} \), \( c_{3} \) and the counterterms \( K_{i} \). The
fitting procedure have shown that \( K_{1}(K_{4}) \) is strongly correlated(anticorrelated)
with \( c_{2}(c_{3}) \). Thus, neglecting small terms of order \( M^{2}_{\pi }/M^{2}_{K} \)
in (\ref{op2deltai12}) and (\ref{op2deltai32}), the coupling constants \( K_{1} \)
and \( K_{4} \) can be safely absorbed by a proper redefinition of the coupling
constants \( c_{2} \) and \( c_{3} \):

\begin{equation}
\label{redefinition}
c_{2}\rightarrow c^{'}_{2}=c_{2}-\frac{2}{9}M^{2}_{K}K_{1}\: ,\; \; c_{3}\rightarrow c^{'}_{3}=c_{3}+\frac{1}{3}M^{2}_{K}K_{4}\: .
\end{equation}
 In this way, the contributions \( A^{\left( 2\right) }_{i} \), \( A^{loop}_{i} \)
and \( A^{st.ct.}_{i} \) are expressed in terms of \( c^{'}_{2} \) and \( c^{'}_{3} \)
while the weak counterterm contributions \( A^{wk.ct.}_{i} \) are determined
by \( K_{2} \), \( K_{3} \), \( K_{5} \), \( K_{6} \) and \( K_{7} \).
The results of the fit are summarized in Table \ref{resulttable} and in the
second column of Table \ref{chptfit}. Obviously, the experimentally measured
isospin amplitudes are in a excellent agreement within the applied chiral scheme.
Moreover, the weak contributions of the quadratic amplitudes \( \zeta _{1} \),
\( \zeta _{3} \), \( \xi _{1} \), \( \xi _{3} \), \( \xi ^{'}_{3} \) are
proportional to the weak contributions of the constant and linear amplitudes
and therefore are independent of the weak couplings \( K_{i} \). Thus, the
good consistency between the measured and predicted values of the quadratic
amplitudes justifies the applicability of the chiral approach at \( O\left( p^{4}\right)  \).

Based on the fitted values\footnote{%
the fit has been repeated by excluding the requirement for the smallness of
the relative isospin phases
} of \( c^{'}_{2} \), \( c^{'}_{3} \) and \( K_{i} \) and taking into account
the imaginary parts of the one-loop contributions from Table \ref{loop}, one
can predict the phases of \( K\rightarrow 3\pi  \) amplitudes. The obtained
phases of the constant and linear amplitudes are given in Table \ref{phases}.
The imaginary parts of the transition amplitudes were expanded in Dalitz plot
variables and calculated numerically. Therefore, considering the values in the
table one have to be aware that the phases of the amplitudes \( \beta  \) include
the kinematically dependent phases of the constant amplitudes \( \alpha  \)
estimated at the lowest order in momentum.

\section{\label{resonance}Comparison with weak resonance models}

The results from the ChPT fit allow to make some remarks of the validity of
weak resonance models. In what follows we will consider weak deformation and
factorization models \cite{Ecker}. Both models rely on the hypothesis that
the strong chiral Lagrangian dominates the features of the \( \Delta S=1 \)
effective lagrangian. Due to the saturation of the strong couplings of \( O\left( p^{4}\right)  \)
by resonance exchange, the considered models can be also treated as models for
the resonance contributions to the weak couplings. As it was shown in \cite{Ecker},
the resonance model leads to a scale independent prediction \( K_{2}-K_{3}\simeq 2.6\times 10^{-9} \),
which seems to be compatible with the values in Table \ref{resulttable}. A
further assumption that only vector resonance contribute implies \( K_{1}=K_{2}=0 \)
and \( K_{3}\neq 0 \). Even taking into account the scale dependence\footnote{%
the scale dependencies of the weak coupling can be found in \cite{Ecker}
} of \( K_{2} \), the vector resonance model is far from describing the data.
Now lets consider the prediction

\begin{equation}
\label{wdmfm}
-K_{1}\simeq K_{2}=K_{3}\simeq k_{f}\left( 1.7\times 10^{-9}\right) \: ,
\end{equation}
which comes from the weak deformation and the factorization models. The scale
factor \( k_{f} \) in (\ref{wdmfm}) is equal to \( 1/2 \) for the weak deformation
model and is around unity for the factorization model. Contrariwise, the obtained
values for the couplings \( K_{2} \) and \( K_{3} \) require \( k_{f}\approx 4 \).
The relatively large uncertainties of the weak coupling constants and their
sizable scale dependence can reduce this estimation by a factor of \( 2 \),
which unfortunately remains sufficiently far from the preferred by the factorization
model value of \( \approx 1 \).
\vspace*{0.5cm}

\section{\label{conclusions}Conclusions}

The applied ChPT approach establish certain relations between \( K\rightarrow 2\pi  \)
amplitudes and the constant and the linear amplitudes in \( K\rightarrow 3\pi  \)
decays. Moreover, the quadratic \( K\rightarrow 3\pi  \) amplitudes are strongly
correlated to the precisely measured constant and linear ones and therefore
are independent of the weak coupling constants. The fit to the experimentally
measured isospin amplitudes have shown that the chiral constraints are well
satisfied not only for the \( \Delta I=1/2 \) amplitudes, but also for the
\( \Delta I=3/2 \) ones. This fact leads to the conclusion of the validity
of the chiral scheme in the description of the nonleptonic kaon decays.

In addition to these achievements, the presented analysis has two main disadvantages.
The first one is that we have omitted the radiative corrections to \( K\rightarrow 3\pi  \)
amplitudes. As it was shown in \cite{Belkov}, these corrections may introduce
significant effects especially in the case of \( K^{+}\rightarrow \pi ^{+}\pi ^{+}\pi ^{-} \)
decay. The second disadvantage is neglecting the strong rescattering phases
in the estimation of \( K\rightarrow 3\pi  \) amplitudes. The latter is due
to the fact that the experimental information in CP conserving \( K\rightarrow 3\pi  \)
decays is insufficient to provide these phases. From the other side, the real
counterterms have been fitted to the assumed real isospin amplitudes and as
a consequence the obtained values of the strong rescattering phases are suppressed.
Therefore, it would be more desirable to fix the phases, for example, in precise
time-interferometry experiments \cite{Ambrosio} and to use the measurements
as an input to the fitting procedure.
\vspace*{0.5cm}

I am grateful to P.Hristov and L.Litov for useful discussions and comments.

\clearpage
\begin{table}[!t]

\caption{\label{expslopes}Experimental data for the widths and the Dalitz plot slopes
in CP conserving \protect\( K\rightarrow 3\pi \protect \) decays.\protect \\
}
{\centering \begin{tabular}{|l|c|c|c|c|}
\hline 
{\small Decay}&
{\small \( \Gamma \left( s^{-1}\right) \times 10^{-6} \)}&
{\small \( g\times 10^{1} \)}&
{\small \( h\times 10^{2} \)}&
{\small \( k\times 10^{2} \)}\\
\hline 
\hline 
{\small \( K_{L}\rightarrow \pi ^{+}\pi ^{-}\pi ^{0} \)}&
{\small \( 2.43\pm 0.04 \)}&
{\small \( 6.78\pm 0.08 \)}&
{\small \( 7.6\pm 0.6 \)}&
{\small \( 0.99\pm 0.15 \)}\\
\hline 
{\small \( K_{L}\rightarrow \pi ^{0}\pi ^{0}\pi ^{0} \)}&
{\small \( 4.08\pm 0.06 \)}&
{\small }&
{\small \( -0.50\pm 0.23 \)}&
{\small }\\
\hline 
{\small \( K^{+}\rightarrow \pi ^{+}\pi ^{+}\pi ^{-} \)}&
{\small \( 4.52\pm 0.04 \)}&
{\small -\( 2.154\pm 0.035 \)}&
{\small \( 1.2\pm 0.8 \)}&
{\small \( -1.01\pm 0.34 \)}\\
\hline 
{\small \( K^{+}\rightarrow \pi ^{0}\pi ^{0}\pi ^{+} \)}&
{\small \( 1.40\pm 0.03 \)}&
{\small \( 6.52\pm 0.31 \)}&
{\small \( 5.7\pm 1.8 \)}&
{\small \( 1.97\pm 0.54 \)}\\
\hline 
{\small \( K_{S}\rightarrow \pi ^{+}\pi ^{-}\pi ^{0} \)}&
{\small \( 0.0036\pm 0.0013 \)}&
{\small }&
{\small }&
\\
\hline 
\end{tabular}\small \par}\end{table}

\begin{table}[!t]

\caption{\label{chptfit}Values of the isospin amplitudes in units of \protect\( 10^{-8}\protect \),
obtained from the phenomenological fit to the experimental data (first column)
and from the fit within \protect\( O\left( p^{4}\right) \protect \) ChPT (second
column).\protect \\
}
{\centering \begin{tabular}{|c|c|c|}
\hline 
&
Experiment&
\( O\left( p^{4}\right)  \)\\
\hline 
\hline 
\( a_{1/2} \), \( keV \)&
\( 0.4665\pm 0.0010 \)&
\( 0.4665 \)\\
\hline 
\( a_{3/2} \), \( keV \)&
\( 0.02116\pm 0.00007 \)&
\( 0.02115 \)\\
\hline 
\( \alpha _{1} \)&
\( 92.12\pm 0.34 \)&
\( 92.11 \)\\
\hline 
\( \alpha _{3} \)&
\( -6.41\pm 0.44 \)&
\( -6.97 \)\\
\hline 
\( \beta _{1} \)&
\( -26.73\pm 0.39 \)&
\( -26.76 \)\\
\hline 
\( \beta _{3} \)&
\( -2.26\pm 0.44 \)&
\( -2.17 \)\\
\hline 
\( \gamma _{3} \)&
\( 2.89\pm 0.28 \)&
\( 2.98 \)\\
\hline 
\( \zeta _{1} \)&
\( -0.38\pm 0.19 \)&
\( -0.51 \)\\
\hline 
\( \zeta _{3} \)&
\( -0.09\pm 0.10 \)&
\( -0.008 \)\\
\hline 
\( \xi _{1} \)&
\( -1.80\pm 0.29 \)&
\( -1.66 \)\\
\hline 
\( \xi _{3} \)&
\( 0.17\pm 0.15 \)&
\( 0.07 \)\\
\hline 
\( \xi ^{'}_{3} \)&
\( -0.57\pm 0.41 \)&
\( -0.15 \)\\
\hline 
\( \chi ^{2}/n.d.f \)&
\( 14.2/5 \)&
\( 13.9/14 \)\\
\hline 
\end{tabular}\par}\end{table}
\clearpage
\begin{table}[!t]

\caption{\label{loop}Values of the reduced one-loop amplitudes \protect\( a^{\left( 8\right) }_{i}\protect \)
and \protect\( a^{\left( 27\right) }_{i}\protect \) for the renormalization
scale \protect\( \mu =M_{\eta }\protect \). \protect \\
}
{\centering \begin{tabular}{|c|c|c|c|c|}
\hline 
&
\( \Re e\: a^{\left( 8\right) }_{i} \)&
\( \Im m\: a^{\left( 8\right) }_{i} \)&
\( \Re e\: a^{\left( 27\right) }_{i} \)&
\( \Im m\: a^{\left( 27\right) }_{i} \)\\
\hline 
\hline 
\( A_{0} \)&
\( 1.79 \)&
\( 2.22 \)&
\( -1.26 \)&
\( -1.48 \)\\
\hline 
\( A_{2} \)&
-&
-&
\( -1.94 \)&
\( 4.63 \)\\
\hline 
\( \alpha _{1} \)&
\( 2.31 \)&
\( 1.00 \)&
\( -1.21 \)&
\( -0.67 \)\\
\hline 
\( \alpha _{3} \)&
-&
-&
\( 22.5 \)&
\( 6.70 \)\\
\hline 
\( \beta _{1} \)&
\( -1.07 \)&
\( 0.50 \)&
\( 0.70 \)&
\( 0.33 \)\\
\hline 
\( \beta _{3} \)&
-&
-&
\( -0.63 \)&
\( -2.19 \)\\
\hline 
\( \gamma _{3} \)&
-&
-&
\( -3.58 \)&
\( 1.12 \)\\
\hline 
\( \zeta _{1} \)&
\( -0.027 \)&
\( 0.019 \)&
\( 0.013 \)&
\( -0.012 \)\\
\hline 
\( \zeta _{3} \)&
-&
-&
\( 0.035 \)&
\( 0.066 \)\\
\hline 
\( \xi _{1} \)&
\( -0.115 \)&
\( 0 \)&
\( 0.088 \)&
\( 0 \)\\
\hline 
\( \xi _{3} \)&
-&
-&
\( -0.126 \)&
\( 0 \)\\
\hline 
\( \xi ^{'}_{3} \)&
-&
-&
\( 0.294 \)&
\( 0.430 \)\\
\hline 
\end{tabular}\par}\end{table}

\begin{table}[!t]

\caption{\label{li}Values of the strong counterterm coupling constants \protect\( L_{i}\protect \)
for the normalization scale \protect\( \mu =M_{\eta }\protect \) in units of
\protect\( 10^{-3}\protect \).\protect \\
}
{\centering \begin{tabular}{|c|c|c|c|c|}
\hline 
\( L_{1} \)&
\( L_{2} \)&
\( L_{3} \)&
\( L_{4} \)&
\( L_{5} \)\\
\hline 
\hline 
\( 0.6\pm 0.3 \)&
\( 1.75\pm 0.3 \)&
\( -3.5\pm 1.1 \)&
\( 0.0\pm 0.5 \)&
\( 2.2\pm 0.5 \)\\
\hline 
\end{tabular}\par}\end{table}
\clearpage
\begin{table}[!t]

\caption{\label{resulttable}Values of the redefined coupling constants \protect\( c^{'}_{2}\protect \),
\protect\( c^{'}_{3}\protect \) and the counterterms \protect\( K_{i}\protect \)
in units of \protect\( 10^{-9}\protect \) obtained from full fit to the experimentally
measured isospin amplitudes. The errors are from the experimentally measured
amplitudes (first column of Table \ref{chptfit}) and from the uncertainties
in the strong couplings \protect\( L_{i}\protect \) (Table \ref{li}), respectively.\protect \\
}
{\centering \begin{tabular}{|c|c|}
\hline 
\( c^{'}_{2}/F^{2}_{\pi } \)&
\( 65.47\pm 0.14 \)\\
\hline 
\( c^{'}_{3}/F^{2}_{\pi } \)&
\( -0.828\pm 0.003 \)\\
\hline 
\( K_{2} \)&
\( 6.9\pm 0.1\pm 2.2 \)\\
\hline 
\( K_{3} \)&
\( 7.0\pm 0.7\pm 2.0 \)\\
\hline 
\( K_{5} \)&
\( -0.016\pm 0.003\pm 0.010 \)\\
\hline 
\( K_{6} \)&
\( -0.12\pm 0.08\pm 0.02 \)\\
\hline 
\( K_{7} \)&
\( -0.17\pm 0.06\pm 0.001 \)\\
\hline 
\end{tabular}\par}\end{table}

\begin{table}[!t]

\caption{\label{phases}The phases of the constant and linear isospin amplitudes.}
{\centering \begin{tabular}{|c|c|c|c|c|c|}
\hline 
&
\( \alpha _{1} \)&
\( \alpha _{3} \)&
\( \beta _{1} \)&
\( \beta _{3} \)&
\( \gamma _{3} \)\\
\hline 
\hline 
phase&
\( 0.07 \)&
\( 0.09 \)&
\( -0.12 \)&
\( -0.09 \)&
\( -0.03 \)\\
\hline 
\end{tabular}\par}\end{table}

\end{document}